\documentclass[10pt,letterpaper]{article}
\usepackage{custom_diego}
\usepackage{transparent}
\usepackage{tabularx}
\usepackage{makecell}
\usepackage{subfig}
\usepackage{graphicx}
\usepackage{soul}
\usepackage[labelfont=bf]{caption}
\usepackage{array}
\usepackage{multicol}
\newcolumntype{L}[1]{>{\raggedright\let\newline\\\arraybackslash\hspace{0pt}}m{#1}}
\newcolumntype{C}[1]{>{\centering\let\newline\\\arraybackslash\hspace{0pt}}m{#1}}
\newcolumntype{R}[1]{>{\raggedleft\let\newline\\\arraybackslash\hspace{0pt}}m{#1}}

\usepackage[cmex10]{amsmath}

\begin{document}
\title{Graphene Based Plasmonic Tunable Low Pass Filters in the THz Band}

\author{
        D. Correas-Serrano$^{1*}$,
       J.~S.~Gomez-Diaz$^2$,
        J.~Perruisseau-Carrier$^3$,
        and A.~Alvarez-Melcon$^1$}

\address{$^1$Departamento de Tecnolog\'ias de la Informaci\'on y las Comunicaciones,\\
Universidad Polit\'ecnica de Cartagena,
Cartagena, Spain}
\address{$^2$Department of Electrical and Computer Engineering,\\ University of Texas at Austin, Austin, TX 78712 USA}
\address{$^3$Adaptive MicroNanoWave Systems, LEMA/Nanolab,\\
        \'Ecole Polytechnique F\'ed\'erale de Lausanne.
        ,Lausanne, Switzerland}

\email{$^*$diego.correas.serrano@gmail.com} 


%
%

\begin{abstract}
We propose the concept, synthesis, analysis, and design of graphene-based plasmonic tunable low-pass filters operating in the THz band. The proposed structure is composed of a graphene strip transferred onto a dielectric and a set of polysilicon DC gating pads located beneath it. This structure implements a stepped impedance low-pass filter for the propagating surface plasmons by adequately controlling the guiding properties of each strip section through graphene's field effect. A synthesis procedure is presented to design filters with desired specifications in terms of cut-off frequency, in-band performance, and rejection characteristics. The electromagnetic modeling of the structure is efficiently performed by combining an electrostatic scaling law to compute the guiding features of each strip section with a transmission line and transfer-matrix framework, approach further validated via full wave simulations.
The performance of the proposed filters is evaluated in practical scenarios, taking into account the presence of the gating bias and the influence of graphene's losses.
These results, together with the high miniaturization
associated with plasmonic propagation, are very promising for the future use and integration of the proposed filters with other graphene and silicon-based elements in innovative THz communication systems.
\end{abstract}

\section{Introduction}



The unique electrical and optical properties of graphene and its ability to support ultra-confined low-loss surface plasmon polaritons (SPPs) are paving the road to the development of all-integrated graphene plasmonic devices at THz frequencies \cite{Bao12,alu13eucap}. These devices may find application in different areas such as chemical and biological remote sensing, high resolution imaging and tomography, time-domain spectroscopy, atmospheric monitoring, and broadband picocellular or intrasatellite communication networks \cite{federici2005thz,rogalski2003infrared,lubecke98,andress2012ultra}. The desired integrated plasmonic structure must perform multiple complex functions, including signal detection, processing, and transmission. Currently, graphene-based plasmonic components ranging from waveguides \cite{Jablan09,Nikitin11,Christensen12, Sebas12_jap}, antennas \cite{Tamagnone12_apl,Filter13,huang2012design,esquius2014sinusoidallyleaky}, reflectarrays \cite{Carrasco13}, amplifiers \cite{Rana08}, switches \cite{sebas13_Switch}, modulators \cite{Sensale-Rodriguez12},
or phase-shifters \cite{Chen13_phase_shifters} to sensors \cite{Arsat09,Kang10} have already been proposed and investigated.

Terahertz systems also require filtering elements to select target frequency bands and reject thermal radiation that may otherwise saturate sensitive detectors \cite{schuster11, boppel11}.
Several types of THz filters can be found in the literature, such as low pass \cite{lee2011terahertz}, high-pass \cite{wu2003terahertz}, band-pass \cite{lu2011second}, and band-stop filters \cite{cunningham2005terahertz}. However, these implementations are based on bulky and heavy quasi-optical components that cannot be tuned electrically.
In addition, planar plasmonic guided filter have been developed using noble metals \cite{huang08,zhu11_lpf,yun2010theoretical}, but they can operate only at optics and infrared frequencies.
Consequently, there is a clear need to develop planar and miniaturized THz filters able to be integrated in future reconfigurable communications and sensing systems.

\begin{figure}[!t]
\centering
\includegraphics[width=0.6\columnwidth]{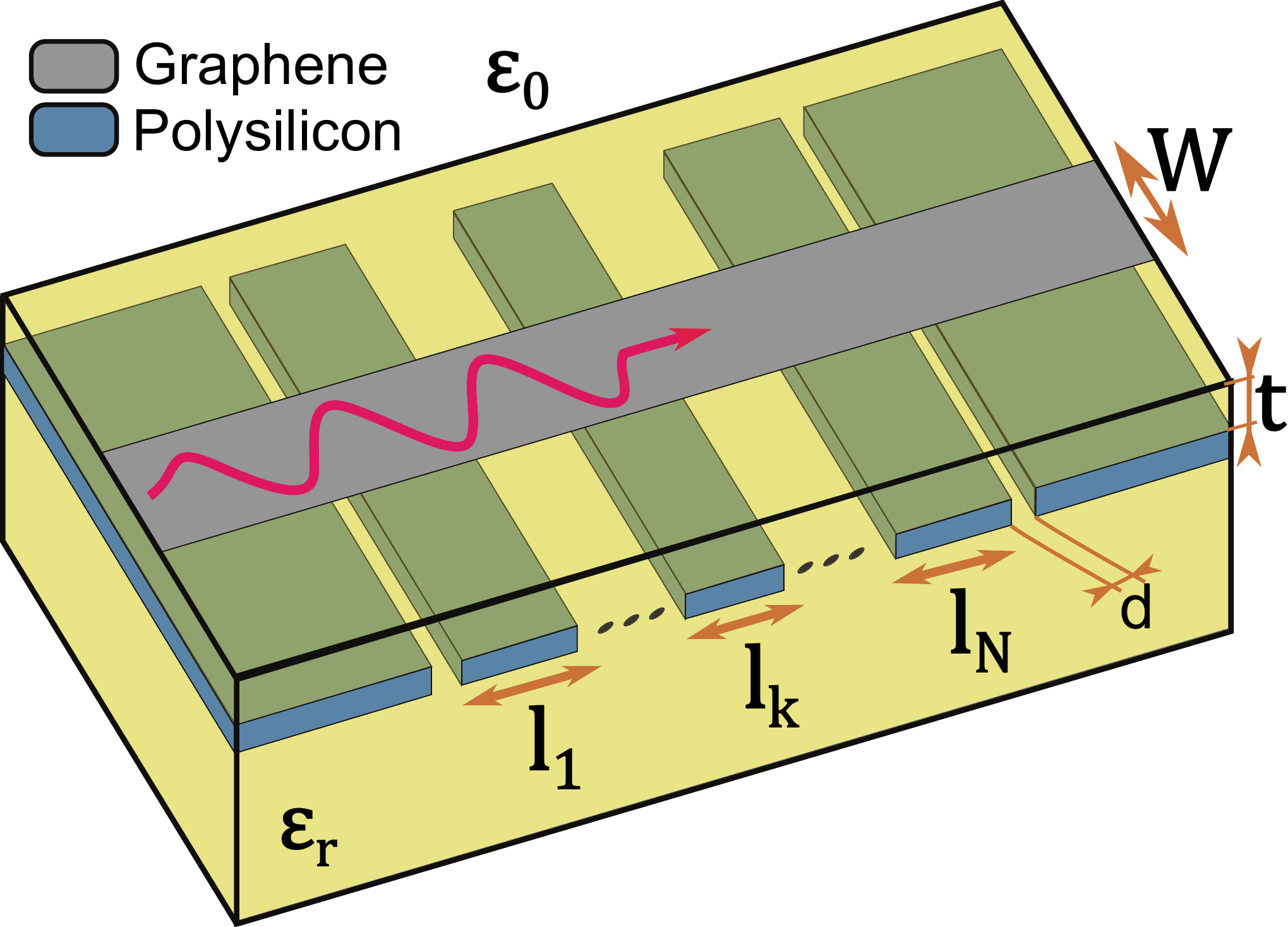}
\caption{Proposed graphene-based THz low pass filter of $N^{th}$ degree. The structure consists of a monolayer graphene strip, with width $W$, and $N$ gating pads located beneath it.}
\label{fig: struct graph}
\end{figure}

In this context, we propose the concept, analysis, and design of graphene-based THz plasmonic reconfigurable low-pass filters. The structure is composed of a graphene strip and several independent polysilicon DC gating pads located beneath it, as depicted in Fig. \ref{fig: struct graph}. The strip supports the propagation of extremely-confined transverse-magnetic (TM) plasmons \cite{Nikitin11, Christensen12} whose guiding characteristics can be dynamically modified along the structure by applying different DC bias voltages to the gating pads. If all pads are equally biased, the structure behaves as a simple plasmonic transmission line (TL) propagating the input waves towards the output port. When a different DC bias is applied to a gating pad, the guiding properties of the strip area located above are modified thanks to graphene's field effect. We apply this concept to implement stepped impedance low pass filters, which are composed of a cascade of transmission lines alternating sections of high and low characteristic impedance. Importantly, the cutoff frequency of the filters can be dynamically tuned by simultaneously modifying the DC bias applied to the gates. A synthesis procedure is then presented to design filters with the desired cutoff frequency, in-band return loss, and rejection characteristics, directly providing the physical length of the gating pads and the required biasing voltages. The electromagnetic modeling of the structure is performed combining a transmission line model with a transfer-matrix approach. To this purpose, a recently introduced graphene electrostatic scaling law \cite{Christensen12} is applied to efficiently compute the propagation constant of the modes supported by the strip as a function of its width, surrounding media, and applied electrostatic DC bias. The proposed approach allows the accurate analysis of the desired filters in just seconds, avoiding large simulation times of general purpose full-wave software.

The fabrication of the proposed filters could be carried out through standard e-beam lithography techniques, and the coupling of power to the  structure may be accomplished through several recently developed techniques for the excitation of SPPs in graphene \cite{esquius2014sinusoidallyleaky,garciadeabajo2013excitation,garciadeabajo2012apex,otto1968excitation,gao2012excitation,peres2013exact,bludov2013excitation,davoyan2012plasmons}.
Rapid advancements are occurring in these areas, and the proposed filters may represent an important step towards innovative THz communication solutions as a key constituting element of future THz plasmonic systems.

In order to illustrate these concepts, several low-pass filters are designed and analyzed, evaluating their performance and reconfiguration capabilities in the THz band. In addition, some practical considerations concerning the implementation of the proposed filters are addressed, discussing in detail the real gating structure and the influence of graphene's losses in the filters performance.

\section{Proposed structure}
\begin{figure}[!t]
\centering
    \includegraphics[width=0.75\columnwidth]{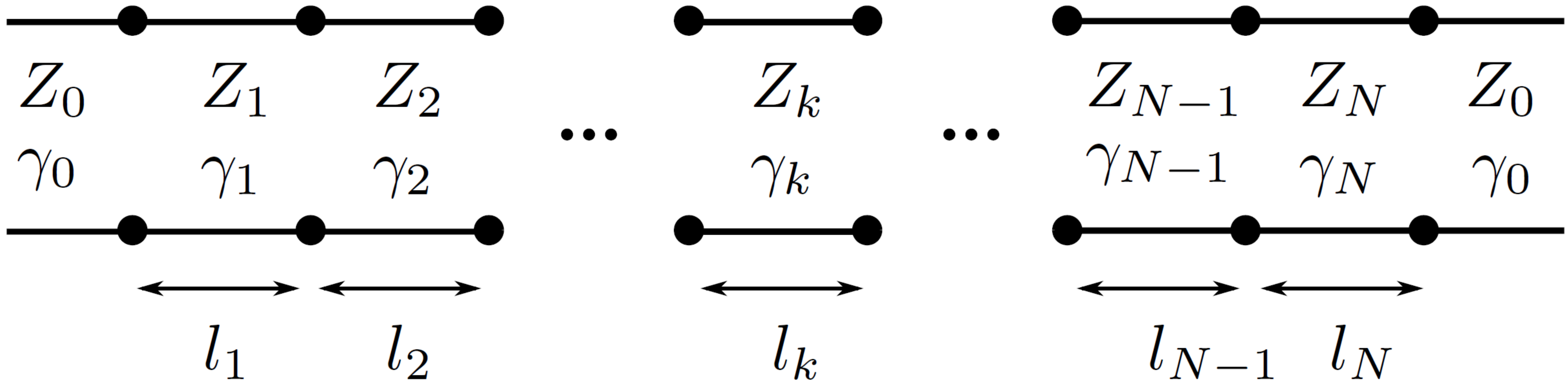}\\
\caption{Equivalent transmission line model of the graphene-based filter shown in Fig~\ref{fig: struct graph}.}
\label{fig: tx line model}
\end{figure}

The proposed structure, depicted in Fig.~\ref{fig: struct graph}, comprises a
graphene ribbon transferred onto a dielectric substrate and a number of polysilicon gating pads beneath the strip. Graphene's surface conductivity ($\sigma$),
like many two dimensional electron gases \cite{burke2000high}, is frequency dependent.
It can be modeled using Kubo's formalism \cite{Hanson08} and below the interband transition threshold can be expressed in closed form as
\begin{equation}\label{eq:conductivity}
\sigma = -j\frac{e^2 k_B T}{\pi \hbar ^2(\omega-j\tau^{-1})} \ln \left\{2\left[1+\cosh\left(\frac{\mu_c}{k_B T}\right)   \right]  \right\},
\end{equation}
where $e$ is the electron charge, $\tau$ is the electron relaxation time in graphene, $k_B$ is Boltzmann's constant, $T$ is temperature, $\omega$ is the angular frequency, and $\mu_c$ is the chemical potential of graphene.
One of the most interesting features of graphene is that its chemical potential can be tuned over a wide range (typically from -1 eV to 1 eV
) by applying a transverse electric field via a DC biased structure, such as the one proposed here. An approximate closed-form expression to relate $\mu_c$ and the applied DC voltage ($V_{DC}$),
is given by
\begin{equation}\label{eq: chemical vdc}
  \centering
  \mu_c \approx \hbar v_F \sqrt{\frac{\pi C_{ox} (V_{DC}-V_{Dirac})}{e}},
\end{equation}
where $V_{Dirac}$ is the voltage at the Dirac point, $C_{ox} \approx \varepsilon_r \varepsilon_0 / t $ is the gate capacitance using the standard parallel-plate approximation,
$\varepsilon_r$ and $t$ are the permittivity and thickness of the gate dielectric, and $v_F$ is the Fermi velocity in graphene ($v_F \approx 10^6$).
In addition, graphene monolayers support the propagation of surface plasmons polaritons
at THz frequencies with moderate losses and extreme confinement.
Several authors have studied the characteristics of the SPPs propagating along graphene ribbons \cite{Nikitin11,Christensen12},
 and transmission line models have been successfully utilized to describe this type of structure \cite{Rana08,sebas13TLmodel,aizin2012transmission}. Using this approach, the structure of Fig. \ref{fig: struct graph} can be modeled as a cascade of transmission lines, as shown in Fig. \ref{fig: tx line model}. The complex-valued characteristic impedance and propagation constant of the transmission lines depend on graphene's chemical potential, and can be largely modified by the DC voltage applied to the gating pads, allowing the synthesis of the proposed filters.

\section{Synthesis and Modeling of \\
Graphene-based Filters}
\subsection{Synthesis Procedure}
The goal of this section is to design a lowpass filter using the structure of Fig.~\ref{fig: struct graph}, with the desired cutoff frequency, rejection characteristics, and inband performance. This structure implements a so-called stepped impedance lowpass filter \cite{Cameron07}, whose equivalent network is presented in Fig.~\ref{fig: tx line model}. As seen, it is composed of the connection of $N$ transmission line sections with lengths $l_k$, propagation constants $\gamma_k$, and characteristic impedances $Z_k$.
The design procedure starts with the calculation of a set of characteristic polynomials able to satisfy the desired specifications in terms of in-band and out-of-band characteristics. Scattering parameters are expressed in terms of these polynomials as follows

\begin{equation}
  S_{21}(\omega) = \frac{1}{\varepsilon E(\omega)} \quad\text{and} \hspace{0.5cm}S_{11} = \frac{F(\omega)}{E(\omega)}.
\end{equation}
The calculation of the $F(\omega)$ and $E(\omega)$ polynomials is done analytically for most useful transfer functions, including Butterworth and generalized Chebyshev responses. Some useful techniques are reported in \cite{Cameron07}. The next step in the design procedure is the election of the electrical length $\theta_c$ of the individual line sections. This parameter directly determines the periodicity of the frequency response when the ideal polynomials are implemented with transmission lines. Smaller values of $\theta_c$ result in a wider spurious-free range, while requiring more abrupt changes in the line impedances. Having decided the value of $\theta_c$, a recursive technique \cite{Cameron07} is applied to extract the normalized values of characteristic impedances ($\bar{Z_k}$). Then, the de-normalization of the calculated characteristic impedances to the real port impedances ($Z_0$) used in the filter implementation is done as $Z_k = \bar{Z}_k Z_0$.

The final and crucial design step consists in finding the design parameters of the physical structure in Fig.~\ref{fig: struct graph}, to implement the prototype circuit at the desired cut-off frequency. To this purpose, the required de-normalized impedance values ($Z_k$) obtained during the above procedure are synthesized using the SPP properties of graphene strips. This can be most efficiently accomplished by appropriately adjusting the chemical potential ($\mu_c$) along the strip through electrostatic gating.
\begin{figure} \centering
\subfloat[]{\label{fig:_Zc_versus_uc}
\includegraphics[width=0.5\columnwidth]{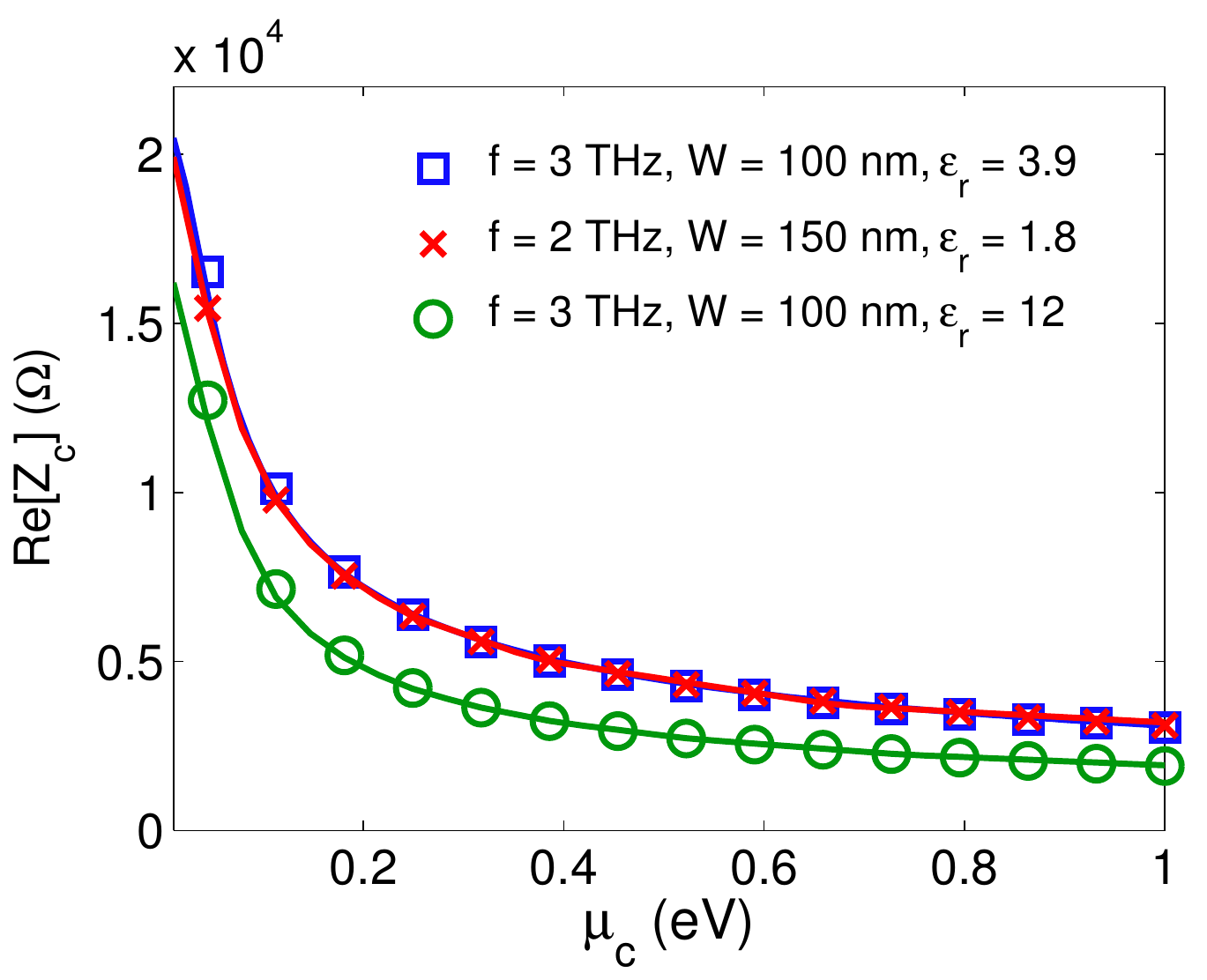}}
\subfloat[]{\label{fig:_beta_versus_uc}
\includegraphics[width=0.5\columnwidth]{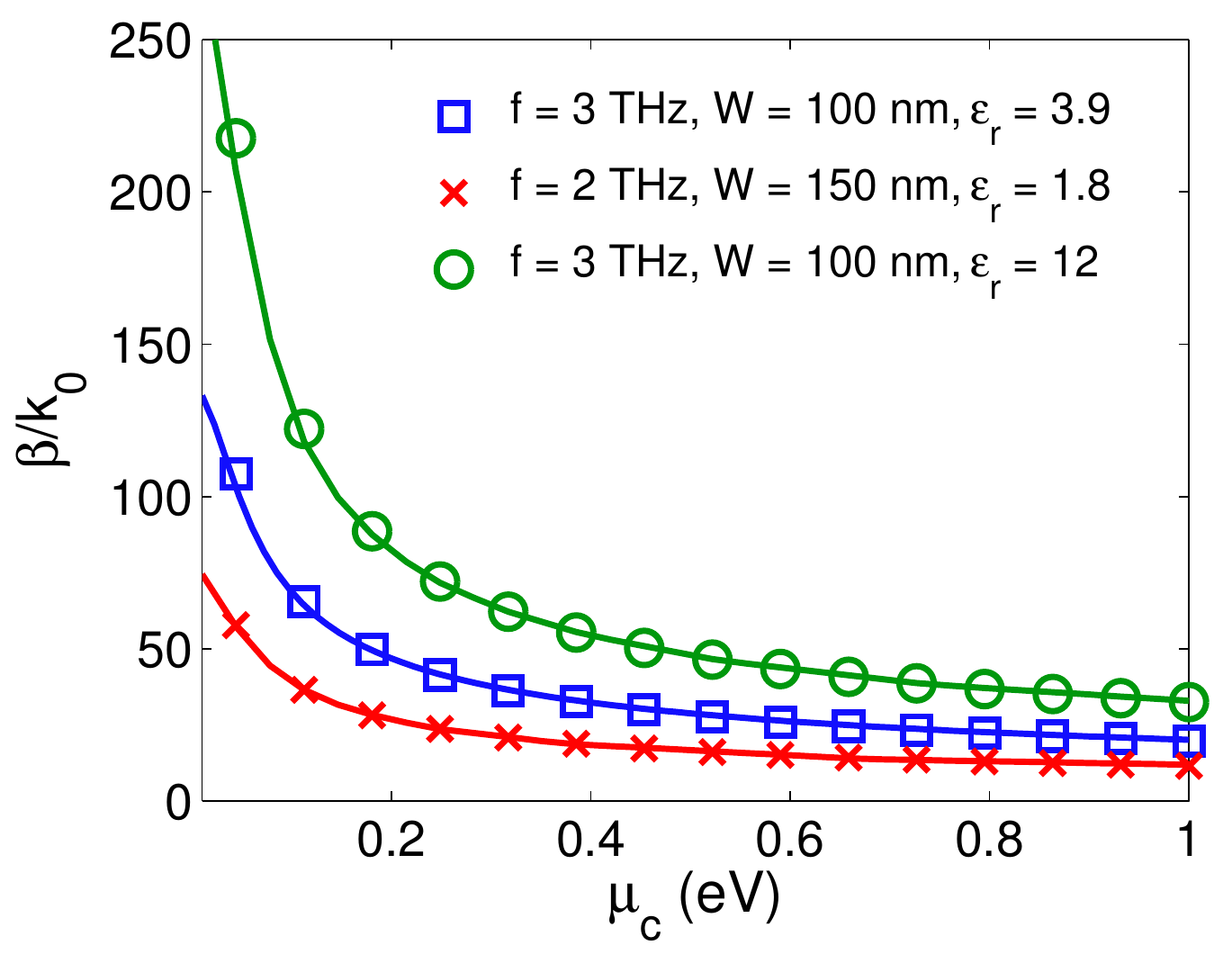}}
\caption{Characteristic impedance (a) and normalized phase constant (b) of the first waveguide-like surface plasmon propagating along different graphene strip configurations versus the chemical potential $\mu_c$. Solid lines have been obtained using the electrostatic approach described in Section III, and markers indicate values computed with HFSS. Graphene parameters are $\tau=1$~ps and $T=300$~K.}
\label{fig:_beta_and_Zc_versus_uc}
\end{figure}
As an illustrative example of this electrical control, we show in Fig.~\ref{fig:_beta_and_Zc_versus_uc} the real part of the characteristic impedance and normalized phase constant of the first waveguide-like mode propagating along a graphene strip, computed versus the chemical potential for graphene strips of different characteristics.
It can be observed in the figure that, for several strip widths and frequencies, the impedances achievable vary from around $20$~k$\Omega$ to several hundreds of ohms for practical chemical potentials in the range $0-1$~eV. Importantly, the calculation of the chemical potentials for all sections
also fixes the values of their propagation constants ($\beta_k$).
This information, combined with the election of $\theta_c$ done during the synthesis phase, allows the calculation of the physical lengths of all graphene strip sections in Fig. \ref{fig: struct graph} using the straightforward relation
\begin{equation}
  l_k=\frac{\theta_c}{\beta_k(\mu_c)}.
\end{equation}
Note that this synthesis procedure is only strictly valid when lossless transmission lines are considered. The presence of losses
in the real structure
leads to small deviations between the actual filter response and the expected synthesized function \cite{Cameron07}, adding some extra round-off inside the filter passband.
Also, the connection of transmission lines of different characteristic impedances is considered to be ideal. In practice, the presence of real gating pads has some influence on the propagation characteristics of the SPP modes propagating along the different graphene strip sections. All these non-ideal effects will be discussed in the last section of the paper.
\subsection{Electromagnetic Modeling}
The electromagnetic modeling of the structure shown in Fig.~\ref{fig: struct graph} is based on the analysis of its equivalent network (see Fig.~\ref{fig: tx line model}) using a transmission line formalism combined with an ABCD transfer-matrix approach \cite{Pozar05}. The main difficulty of this approach is to accurately compute the complex propagation constant and characteristic impedance of the surface plasmons propagating along each section of the strip. The properties and dispersion relation of these plasmons have been studied by several authors \cite{Nikitin11,Christensen12}. Unfortunately, the desired dispersion relation does not admit an analytical solution and one has to resort to full-wave numerical techniques.
Consequently, the use of a transmission line model would generally require the  numerical analysis of the propagating modes along multiple isolated graphene strips of varying characteristics, which remains a computationally costly task. This shortcoming, however, can be elegantly overcome by making use of the quasi-electrostatic nature of surface plasmons in narrow graphene strips, enabling an extremely efficient design and analysis tool. To this end, the scaling law proposed in \cite{Christensen12}, based in the quasi-electrostatic nature of surface plasmons in graphene strips,
 will be used. This approach, which assumes that the strip width is much smaller than the wavelength, establishes that plasmon properties are solely determined by the strip width ($W$), surrounding media ($\varepsilon_r$), and graphene conductivity ($\sigma$).
 Then, once the propagating features of a given plasmonic mode have been obtained, they can be scaled to any arbitrary strip
 by using the scaling parameter

\begin{figure}[t!]
\centering
    \includegraphics[width=0.5\columnwidth]{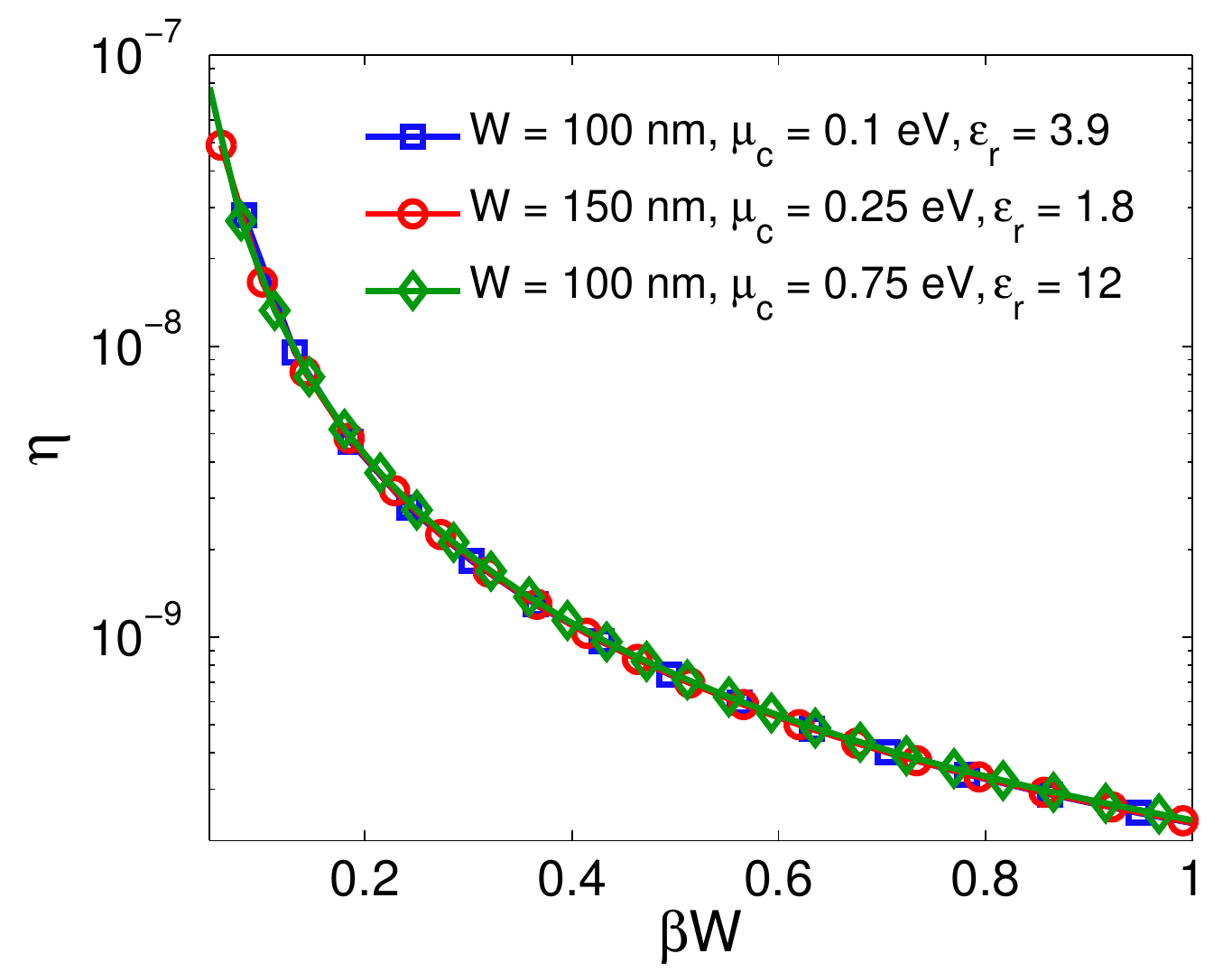}\\
\caption{Scaling parameter $\eta$ versus the product of the phase constant and the strip width ($\beta W$) for different graphene strip configurations. Graphene parameters are $\tau=1$~ps and $T=300$~K.}
\label{fig: eta vs bw}
\end{figure}

\begin{equation}\label{eq: eta}
  \centering
  \eta (\beta, W) = \frac{\mbox{Im}[\sigma(f_\beta)]}{f_\beta W \varepsilon_{eff}},
\end{equation}
 where $f_\beta$ is the frequency where the surface plasmon propagates with a phase constant $\beta$, and $\varepsilon_{eff}=(1+\varepsilon_r)/2$ models the dielectric media. Note that, due to the electrostatic approach employed to derive Eq.~\ref{eq: eta} (see \cite{Christensen12}), the scaling parameter $\eta$ is independent of the operation frequency.

 \begin{table*}[t!]
\renewcommand{\arraystretch}{1.4}
\caption{Design parameters of the first example: a 7th degree filter.}
\label{tab: parameters 182}
\centering
\centering
\begin{tabular}{C{1.5cm}C{1.5cm}C{1.5cm}C{2.2cm}C{2cm}C{2cm}}
\hline\hline
Section      &  $\bar{Z}$      & $l$ (nm) &  $ \mu_{c,nominal}$ (eV)  & $\mu_{c,tuned 1}$ (eV) &   $\mu_{c,tuned 2}$ (eV) \\
\hline
Ports &   $1$          & $500$   &             $0.17$    & $0.27$  &  $0.41$    \\
1,7  & $1.37$       & $382$ &          $0.1$     & $0.15$  &  $0.23$    \\
2,6     &   $0.57$         & $929$  &  $0.51$    & $0.74$  &  $1$    \\
3,5   &   $2.26$        & $232$   &    $0.026$   & $0.06$  &  $0.1$    \\
4  &   $0.45$         & $1172$    &     $0.79$  & $1$  &  $1$    \\
\hline\hline
\end{tabular}
\end{table*}

\begin{figure*} \centering
\subfloat[]{\label{fig: result strip matlab hfss 182}
\includegraphics[width=0.5\columnwidth]{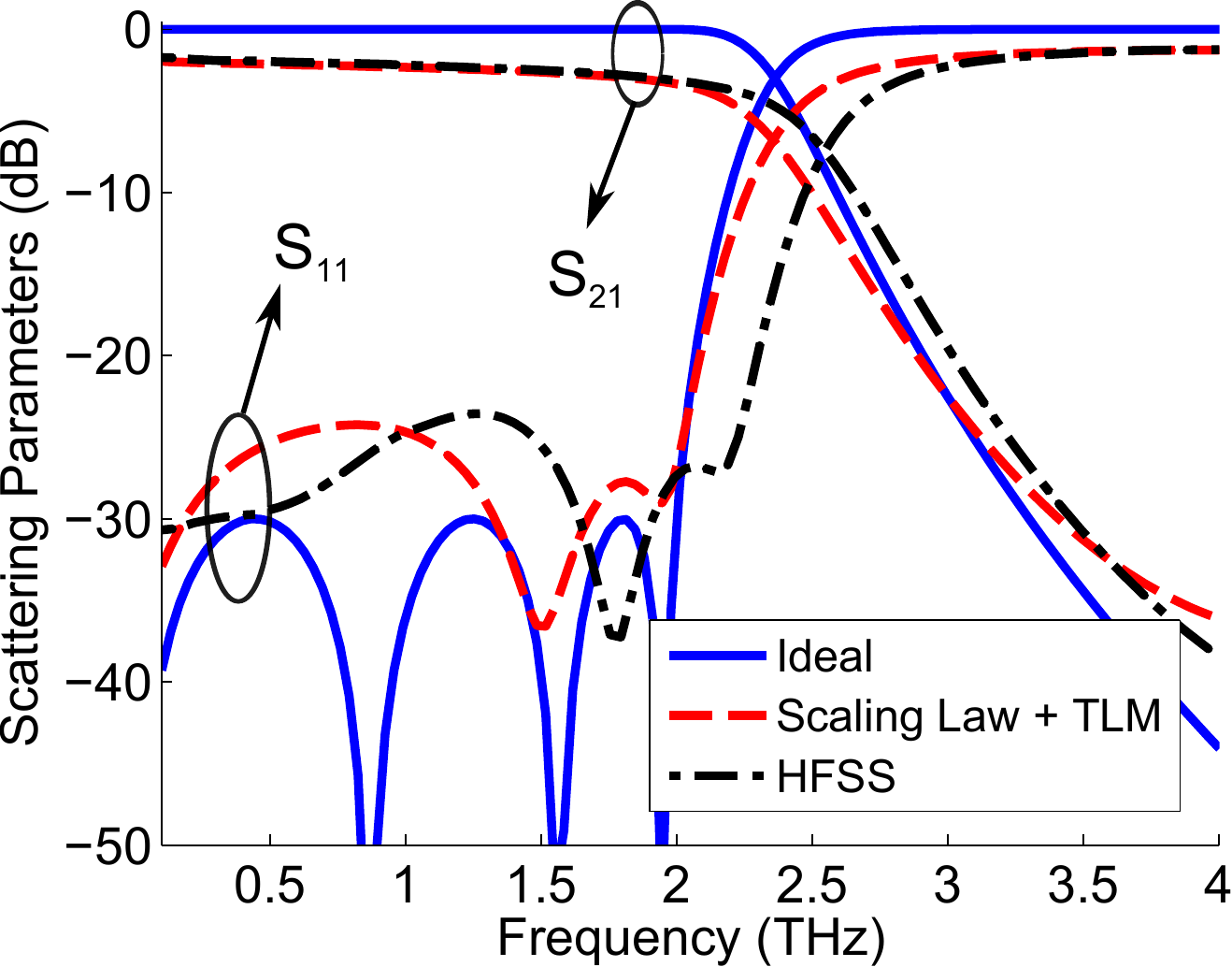}}
\subfloat[]{\label{fig: tunability 182}
\includegraphics[width=0.5\columnwidth]{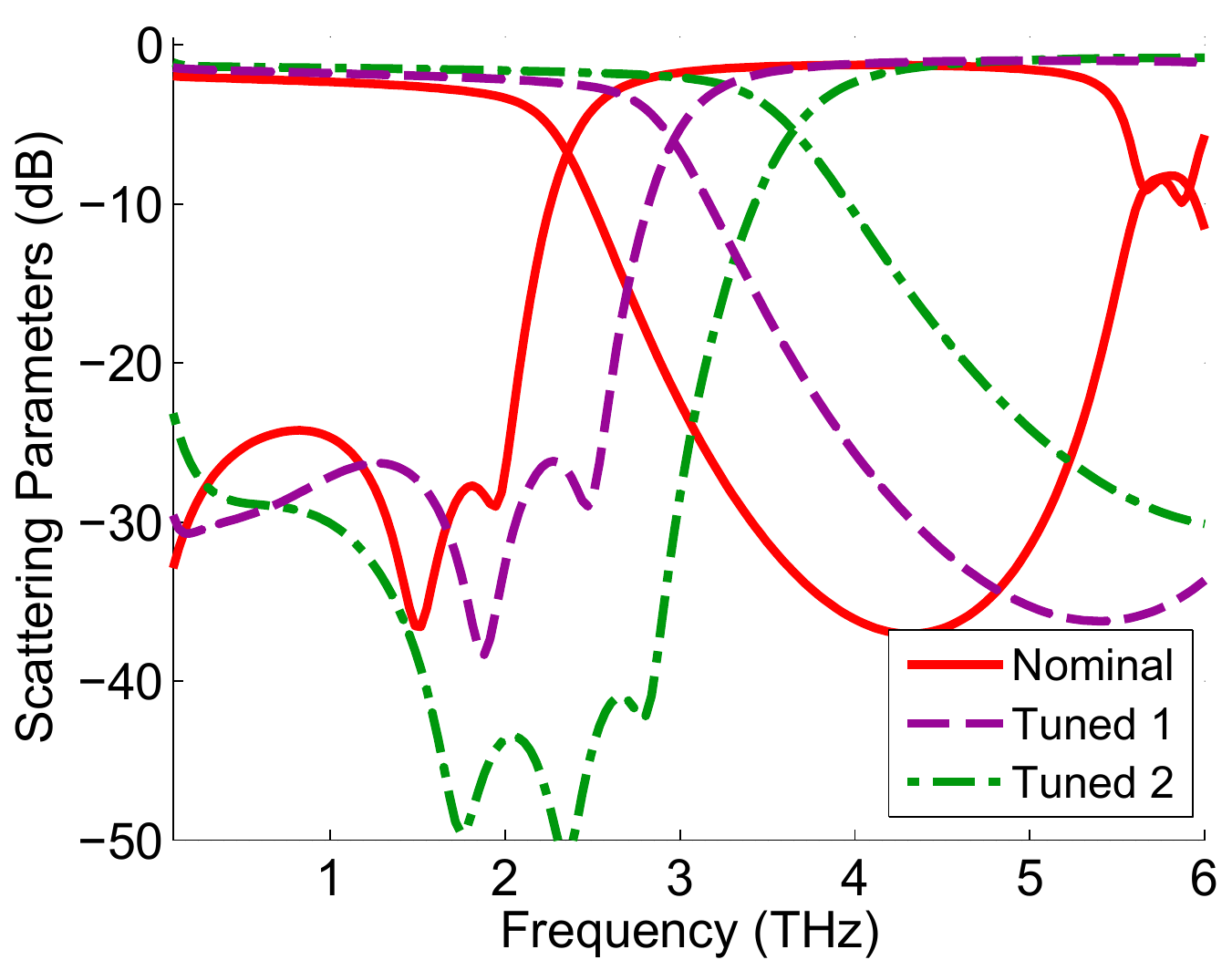}}
\caption{Scattering parameters of a $7^{th}$ degree filter implemented using the structure depicted in Fig.~\ref{fig: struct graph}. The design parameters are shown in Table~\ref{tab: parameters 182}, the strip width is $150$~nm, and a dielectric of $\varepsilon_r = 1.8$ is employed as a substrate. (a) Nominal filter designed to have a cutoff frequency of $2.3$~THz. Results are obtained using the ideal synthesis procedure, the transmission line approach combined with the scaling law, and the commercial software HFSS. (b) Reconfiguration possibilities of the filter obtained by adequately controlling the DC bias of the different gating pads.} \label{fig:_results 182}
\end{figure*}
This scaling law is applied to efficiently compute the SPP propagating features along any graphene strip. The process is as follows. First, the scaling parameter $\eta$ related to the desired surface mode is obtained
using a \emph{single} full-wave simulation.
Note that this simulation is performed only once, and the $\eta - \beta W$ curve computed
will be employed for the design of any filter, regardless of the strip width, graphene conductivity, or supporting substrate.
Fig. \ref{fig: eta vs bw} confirms that $\eta$ only depends on the product $\beta W$ \cite{Christensen12} and that surface plasmons propagating on different strips lead to
the same scaling parameter.
Second, the phase constants of surface plasmons propagating along strips of arbitrary characteristics are computed using $\eta$. To this purpose, the scaling parameter is computed at the desired operation frequency $f_\beta$ and strip width $W$ using Eq.~\ref{eq: eta}, and the corresponding value of $\beta$ is then retrieved using the information of Fig.~\ref{fig: eta vs bw}.
 Finally, the attenuation constant is found as $\alpha = 1/(2 L_p)$, where $L_p$ is the $1/e$ decay distance of the power. The propagation distance of plasmons ($L_p$) is mainly controlled by the electron relaxation time of graphene $\tau$, and it can be approximately obtained by $v_g \tau$, where $v_g$ is the plasmon's group velocity ($v_g = \mbox{d}\omega_p/\mbox{d}\beta$) \cite{Christensen12}. Therefore, the attenuation constant may be expressed as
 \begin{table*}[!t]
\renewcommand{\arraystretch}{1.4}
\caption{Design parameters of the second example: a 9th degree filter.}
\label{tab: parameters 39}
\centering
\begin{tabular}{C{1.5cm}C{1.5cm}C{1.5cm}C{2.2cm}C{2cm}C{2cm}}

\hline\hline
Section      &  $\bar{Z}$      & $l$ (nm) &  $ \mu_{c,nominal}$ (eV)  & $\mu_{c,tuned 1}$ (eV) &   $\mu_{c,tuned 2}$ (eV) \\
\hline
Ports  &  $1$    &       $200$ & $0.17$        & $0.35$  &  $0.46$    \\
1,9  &  $1.36$  &      $156$  & $0.1$ & $0.21$  &  $0.274$    \\
2,8     & $0.58$         & $367$   &  $0.43$   & $0.93$  &  $1$    \\
3,7   & $2.24$   & $94$  &    $0.035$ & $0.01$  &  $0.12$    \\
4,6  & $0.44$   & $483$  &      $0.69$  & $1$  &  $1$    \\
5  &  $2.48$  & $85$  &        $0.023$  & $0.078$  &  $0.1$    \\
\hline\hline
\end{tabular}
\end{table*}

\begin{figure*}[!h] \centering
\subfloat[]{\label{fig: result2 strip matlab hfss 182}
\includegraphics[width=0.5\columnwidth]{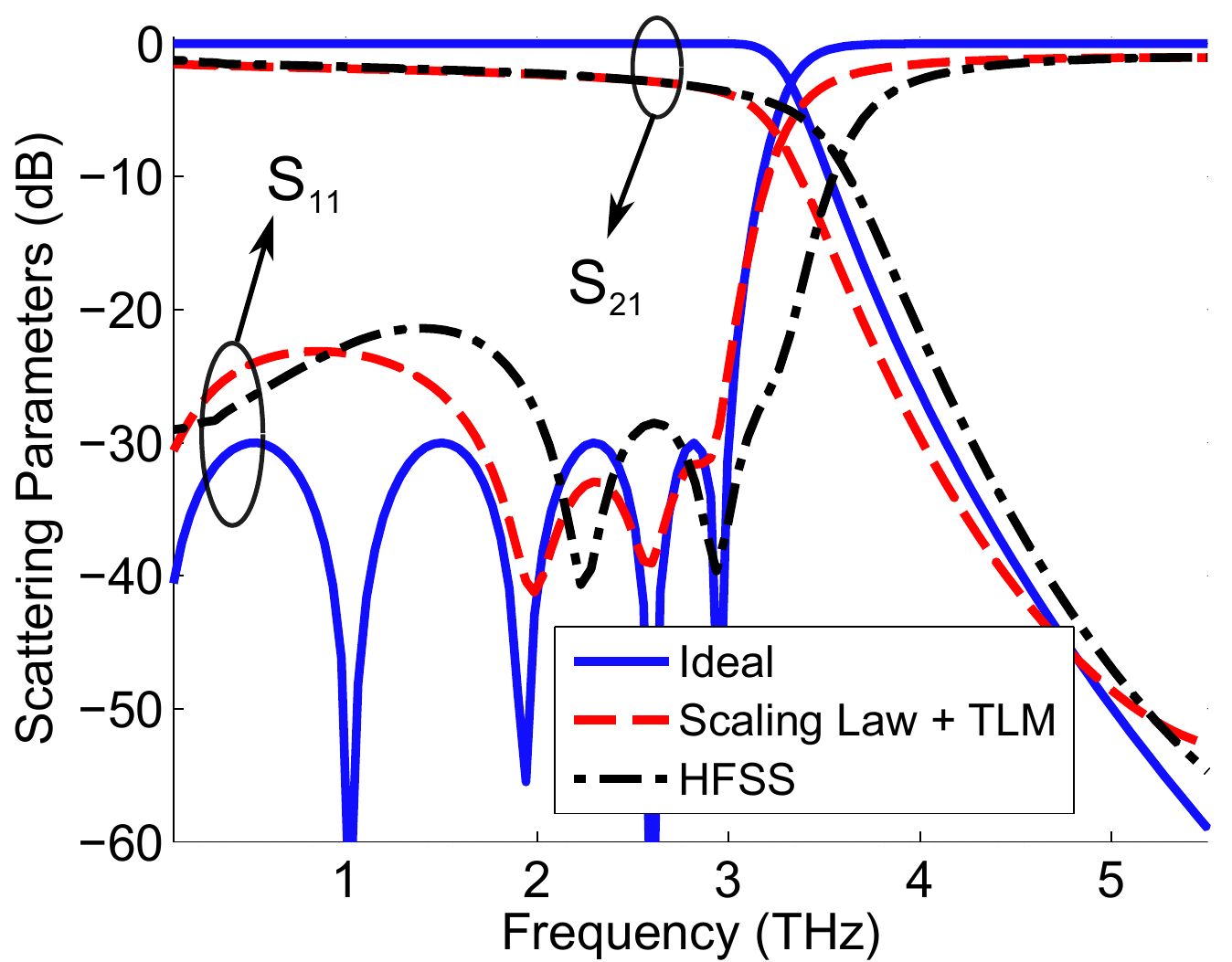}}
\subfloat[]{\label{fig: tunability2 182}
\includegraphics[width=0.5\columnwidth]{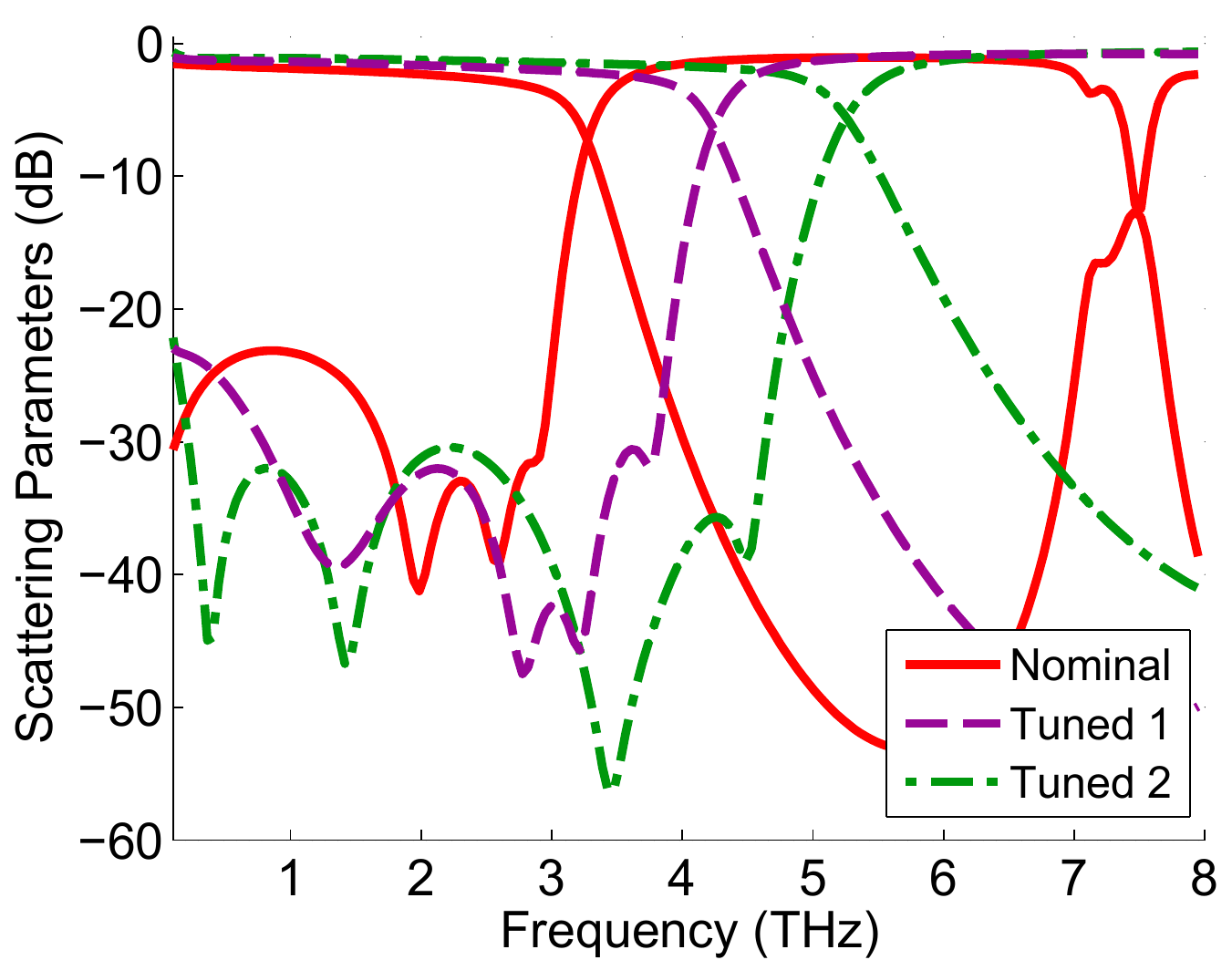}}
\caption{Scattering parameters of a $9^{th}$ degree filter implemented using the structure depicted in Fig.~\ref{fig: struct graph}. The design parameters are shown in Table~\ref{tab: parameters 39}, the strip width is $100$~nm, and a dielectric of $\varepsilon_r = 3.9$ is used as substrate. (a) Nominal filter designed to have a cutoff frequency of $3.3$~THz. Results are obtained using the ideal synthesis procedure, the transmission line approach combined with the scaling law, and the commercial software HFSS. (b) Reconfiguration possibilities of the filter obtained by adequately controlling the DC bias of the different gating pads.} \label{fig:_results 39}
\end{figure*}
\begin{equation}\label{eq: alpha}
  \centering
  \alpha = \frac{1}{2 v_g \tau}.
\end{equation}
Once the complex propagation constant $\gamma = \alpha + j\beta$ of the propagating TM plasmon is known, its characteristic impedance is obtained as \cite{Pozar05}
\begin{equation}\label{eq: Zc}
  \centering
  Z_C = \frac{\gamma}{j \omega \varepsilon_0 \varepsilon_{eff}}.
\end{equation}

The accuracy of this approach is validated in Fig. \ref{fig:_beta_and_Zc_versus_uc}, where the characteristic impedance and phase constant of plasmons propagating along different strip configurations are computed using both the proposed technique and the finite element method (FEM) software Ansoft HFSS. There, the graphene strip is modeled as an infinitesimally thin layer where surface impedance boundary conditions are imposed ($Z_{surf} = 1/\sigma$).

The combination of the graphene strip scaling law with a transmission line and transfer-matrix approach permits an extremely fast electromagnetic analysis of the proposed filtering structure (see Fig.~\ref{fig: struct graph}), allowing an efficient implementation of the synthesis technique described in the previous section.

\section{Design Examples}

In this section we design and analyze two low-pass filters
implemented using the structure shown Fig.~\ref{fig: struct graph}. For the sake of generality, the filters have been designed to have different order and cutoff frequencies, considering various strip widths and dielectrics. The performance of the filters is presented in terms of their scattering parameters, referred to the characteristic impedance of the graphene sections at the input and output ports. A comparison between the transmission line model combined with the scaling law and full-wave results using HFSS is shown for both cases, validating the accuracy of the proposed electromagnetic modeling. In addition, the reconfiguration capabilities of the filters are investigated in detail. In this study, we consider a temperature of $T = 300$~K and
graphene with a relaxation time $\tau$ of $1$~ps,
which corresponds approximately to a carrier mobility of $50000$~$cm^2/(V s)$ for a chemical potential of $\mu_c = 0.2$~eV.
Here we focus on the filtering performance of the proposed structures, whereas other practical considerations, such as the presence of the gating pads and their effect on the filter performance, or the influence of losses, will be discussed in the next section.

In the first example, we consider a graphene strip of $150$~nm transferred onto a dielectric of relative permittivity $1.8$. Using this host waveguide, we have designed a $7^{th}$ degree filter with a cutoff frequency of $2.3$ THz. The Chebyshev polynomials were computed using standard techniques \cite{Cameron07} for a maximum theoretical return loss of $30$ dB, and the electrical length $\theta_c$ at the cutoff frequency was set to $37^\circ$. These values were chosen to yield practical values of chemical potential through the synthesis procedure explained in the previous section. The final design parameters of the filter are shown in Table~\ref{tab: parameters 182}. Note that filters of this degree with better roll-off characteristics could be synthesized, but this would further increase the required range of chemical potential values achievable in the structure.
Fig.~\ref{fig: result strip matlab hfss 182} shows the frequency response of the filter, computed using the proposed transmission line approach and the full-wave commercial software HFSS. A good degree of agreement is observed, with very similar attenuation profile and average level of return loss in the passband.
Moreover, the structure presents a low insertion loss level, around $3$~dB, which is a remarkable value at this frequency range.
The slight difference in the cutoff frequency between both approaches is due to the mono-modal nature of the transmission line approach, which neglects
higher order effects at the connection between two adjacent graphene strip sections.
This results in small modifications
in the phase condition of the circuit, which is now fulfilled at a slightly different cutoff frequency. Interestingly, this effect appears to be uniform along the structure, allowing to adjust the filter's cutoff frequency by a small overall scaling of the section lengths. Fig.~\ref{fig: tunability 182} illustrates the reconfiguration possibilities of the designed filter. By adequately controlling the DC voltage applied to the gating pads, following the synthesis procedure described in Section III, the overall electrical length of the device can be modified thus leading to an electric control of the filter's cutoff frequency.
For the sake of clarity, only two additional possible reconfiguration states are shown in the figure (their corresponding design parameters are shown in Table \ref{tab: parameters 182}). However, a continuous range of cutoff frequencies can be easily synthesized. Importantly, the proposed filter allows a dynamic control of the cutoff frequency over $50\%$. This value is mainly limited by two factors. First, the fixed physical length of the gating pads imposes a limit to the maximum frequency shift attainable while maintaining an acceptable variation of the attenuation profile and return loss. Second, the values of the chemical potential employed are limited, by technological reasons \cite{fan95}, to the range of $0 - 1$~eV.

\begin{figure*}[!t]\centering
\subfloat[]{\label{fig: muc and filters a}
\includegraphics[width=0.5\columnwidth]{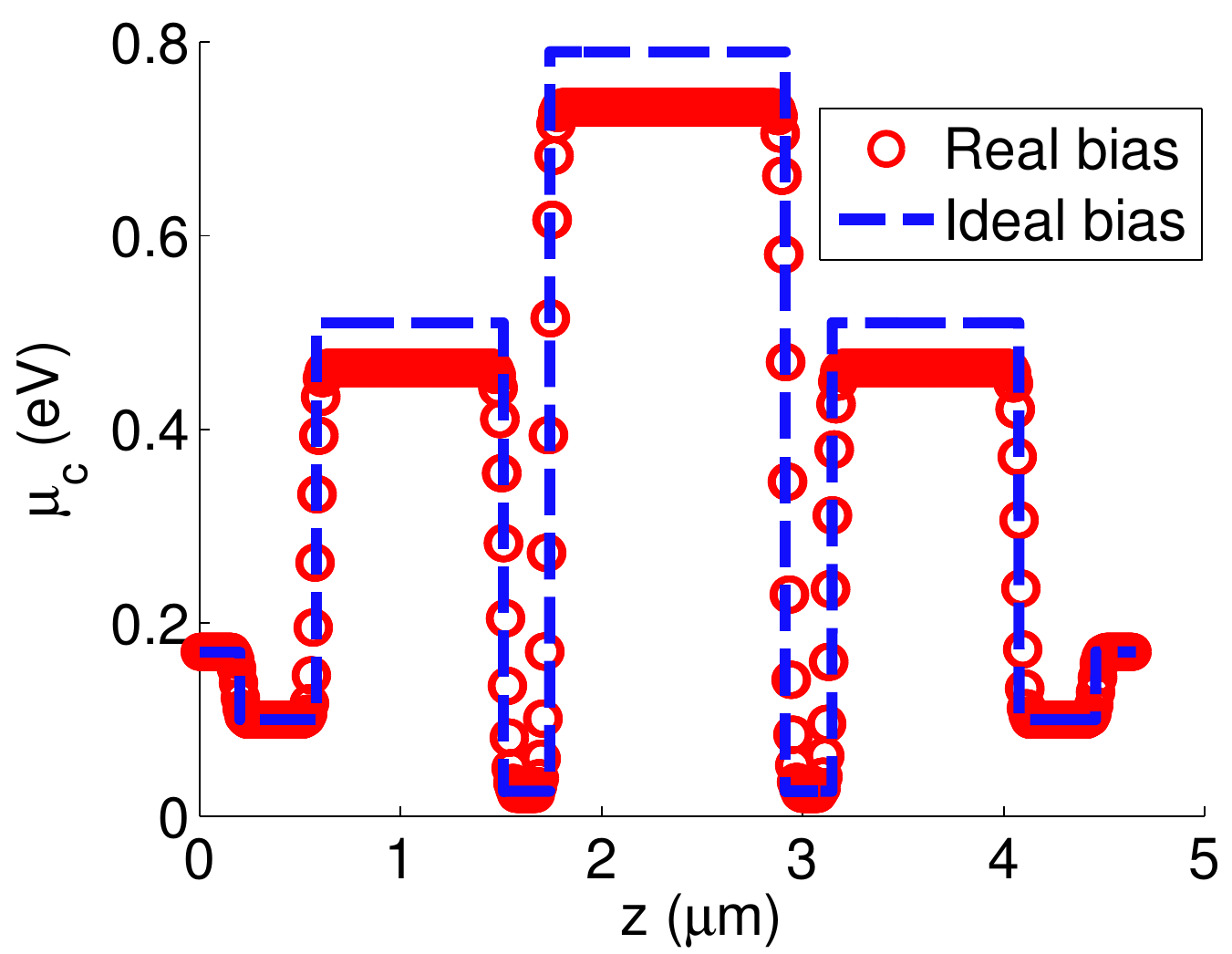}}
\subfloat[]{\label{fig: muc and filters b}
\includegraphics[width=0.5\columnwidth]{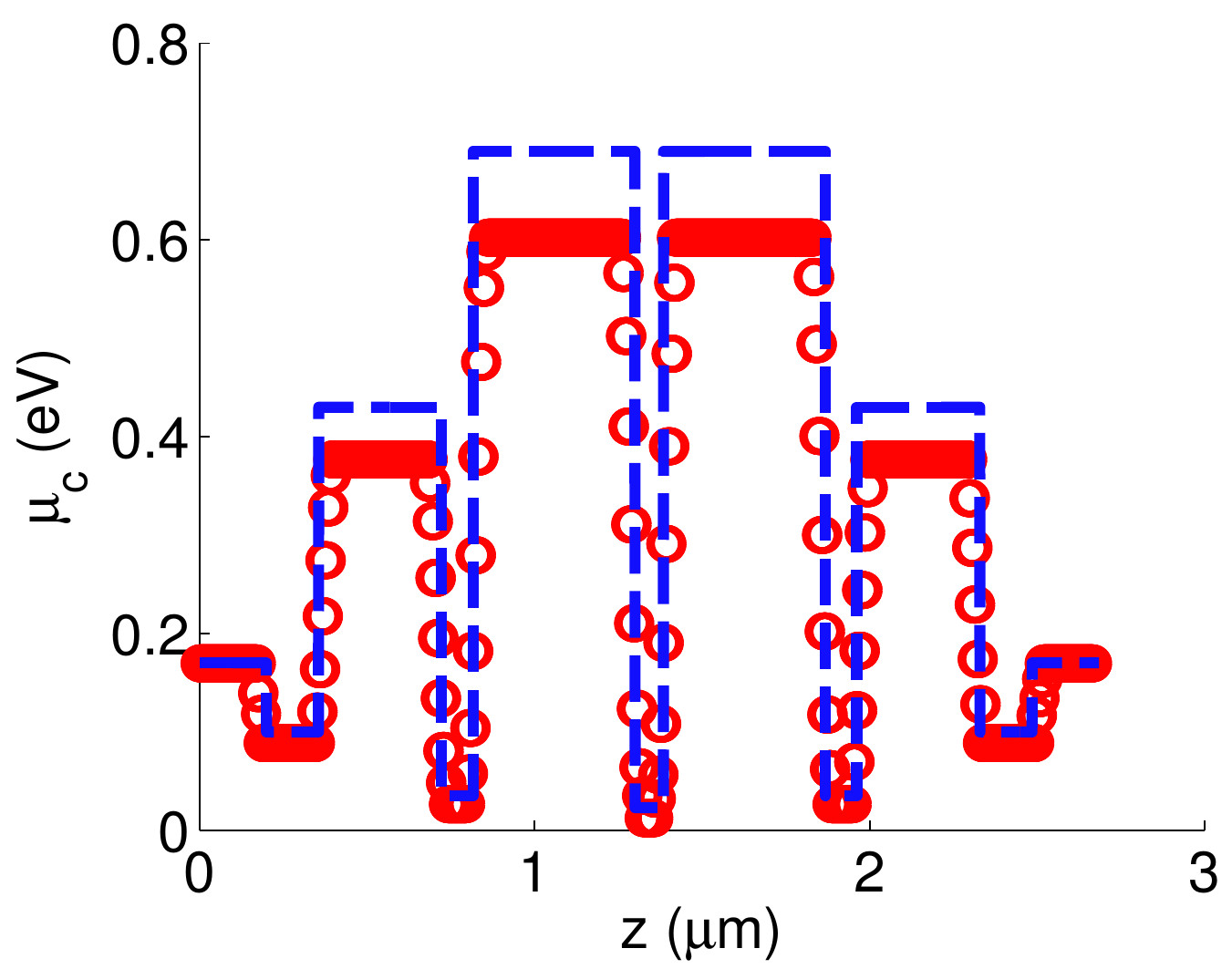}}\\
\subfloat[]{\label{fig: muc and filters c}
\includegraphics[width=0.5\columnwidth]{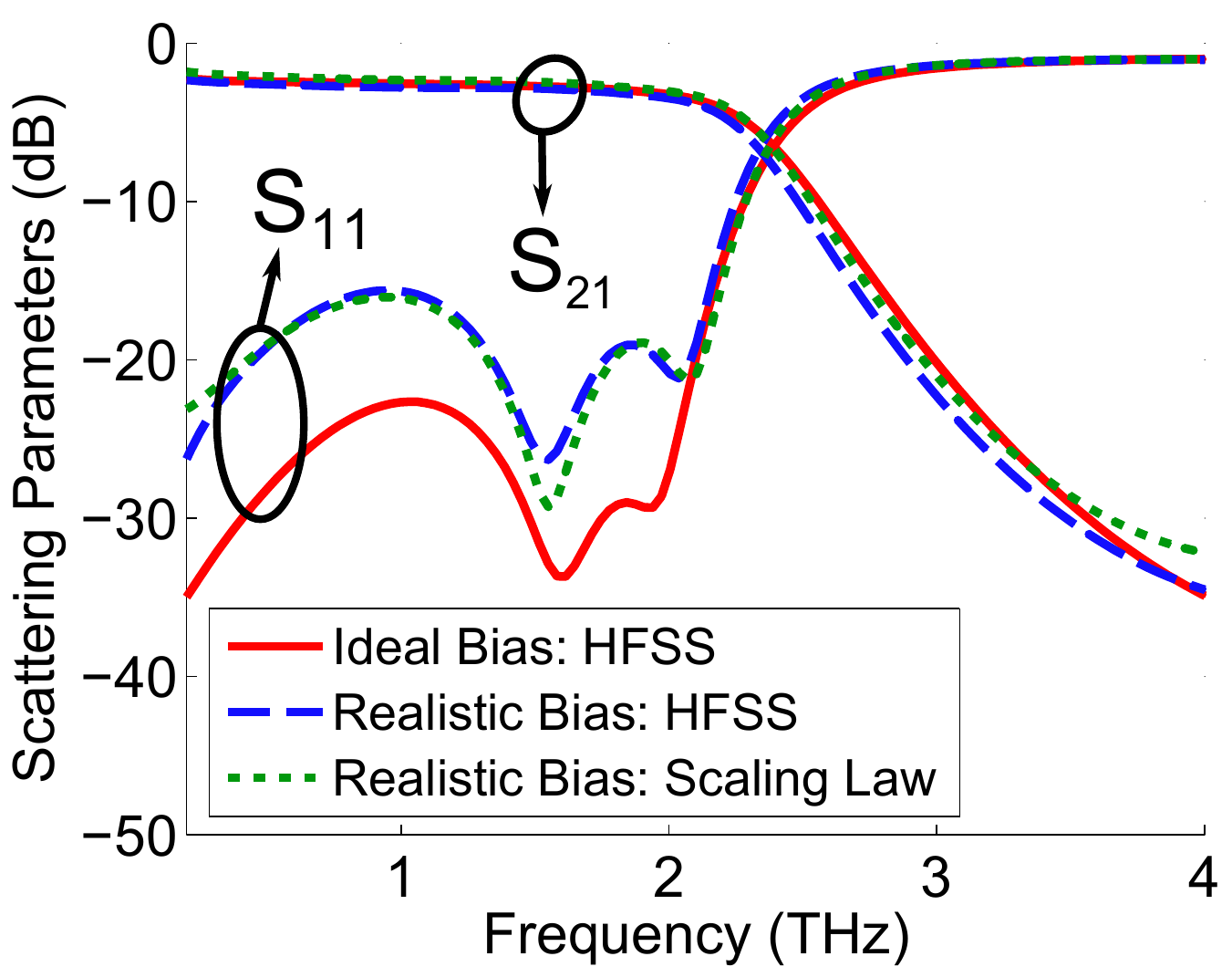}}
\subfloat[]{\label{fig: muc and filters d}
\includegraphics[width=0.5\columnwidth]{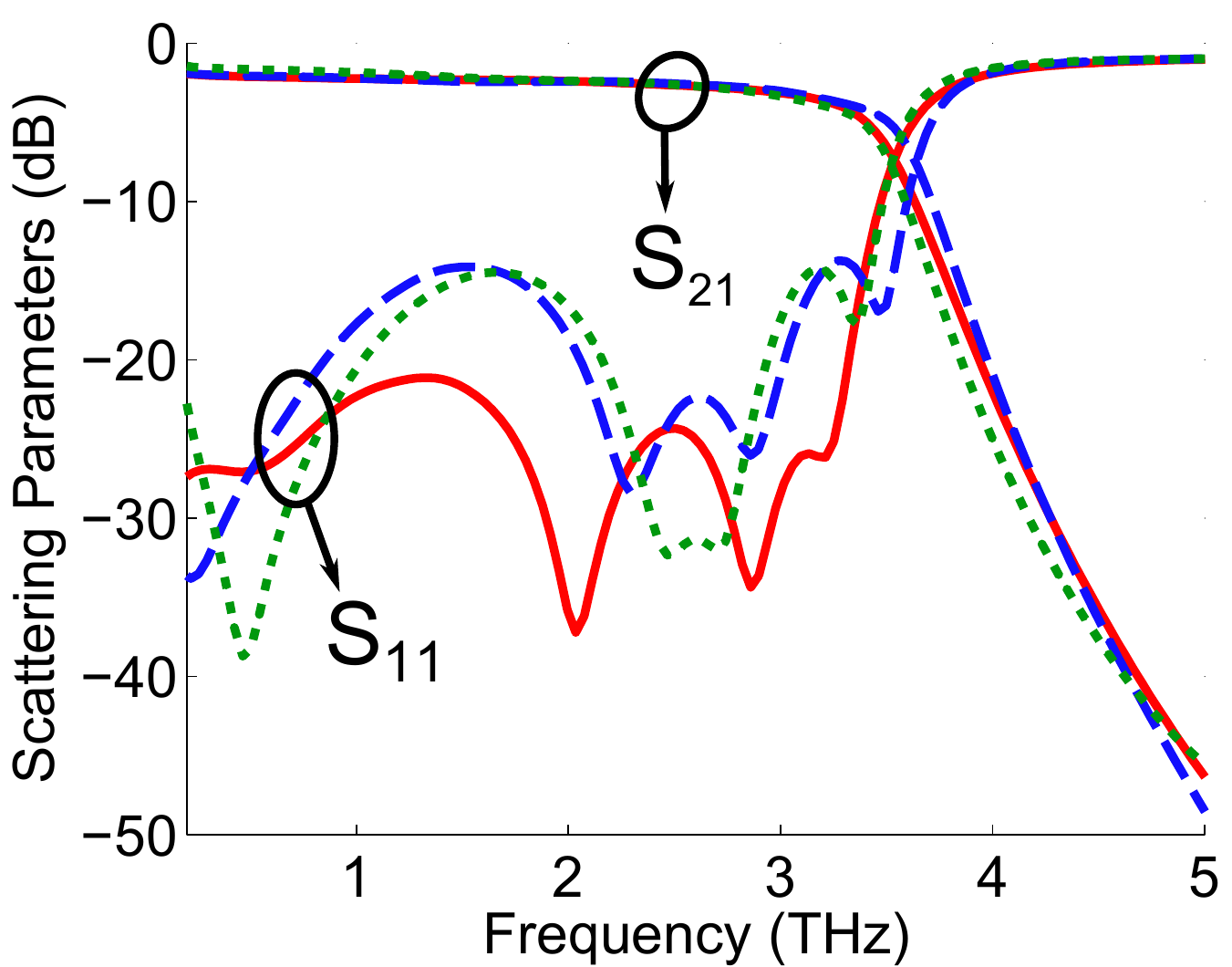}}
\caption{Comparison between ideal and realistic longitudinal chemical potential distributions of the (a) $7^{th}$ and (b) $9^{th}$ degree filters. (c) and (d) show the frequency response of the resulting filters computed using the transmission line approach combined with the scaling law, and the commercial software HFSS.}
\label{fig: muc and filters}
\end{figure*}

\begin{table}[]
\renewcommand{\arraystretch}{1.4}
\caption{Gate Voltages for the $7^{th}$ degree filter}
\label{tab: pads 1}
\centering
\begin{tabular}{cccccc}

\hline\hline
Section   &   ports &  1,7  & 2,6 & 3,5  & 4  \\
\hline
Initial Voltage (V) &  $5.4$  &      $1.9$  & $48.4$ & $0.13$  &  $116$ \\
\hline
Optimized Voltage (V) & $5.4$ &  $1.8$ & $40$ & $0.1$  &  $90$ \\
\hline\hline
\end{tabular}
\end{table}
\begin{table}[]
\renewcommand{\arraystretch}{1.4}
\caption{Gate Voltages for the $9^{th}$ degree filter}
\label{tab: pads 2}
\centering
\begin{tabular}{ccccccc}

\hline\hline
Section   &   ports &  1,9  & 2,8 & 3,7 & 4,6 &5  \\
\hline
Initial Voltage (V) & $2.5$  &  $0.86$ & $15.9$   & $0.11$  & $41$ & $0.05$   \\
\hline
Optimized Voltage (V) & $2.5$  &  $0.68$ & $12$   & $0.06$  & $31$ & $0.05$   \\
\hline\hline
\end{tabular}
\end{table}
The second example is composed of a strip of width $W=100$~nm transferred onto a quartz substrate of $\varepsilon_r = 3.9$. The filter's degree has been increased to $N=9$ and the cutoff frequency is set to $3.3$ THz, with $\theta_c = 39^\circ$. The complete design parameters of the filter are shown in Table \ref{tab: parameters 39} and the frequency response is plotted in Fig.~\ref{fig: result2 strip matlab hfss 182}. Very similar conclusions compared to the previous design can be drawn from the filter response. Importantly, the small difference in the cutoff frequency between the two electromagnetic analysis is also of similar relative magnitude to that of the previous example. Finally, the tunable features of this filter are shown in Fig.~\ref{fig: tunability2 182}, where around $50\%$ of cutoff frequency dynamic control is again obtained.

\section{Practical Considerations}

This section briefly discusses several technological aspects related to the possible practical implementation of the proposed filters, such as the strip biasing and the influence of graphene losses.

\subsection{Rigourous electrostatic biasing of graphene strips}
The results presented in previous section assumed, as a first approximation, an ideal carrier distribution along the graphene strip.
However, this distribution requires strong discontinuities between adjacent sections. In practice, this carrier density profile cannot be achieved because i) there is a physical space between two adjacent gating pads (see Fig. \ref{fig: struct graph}), and ii)
there are fringing effects at the edges of the gating pads, which may modify the carrier density profile.
Here, we show that
good performance is maintained
after rigourously considering these effects.

The analysis of the filter is performed in two different steps. First, a electrostatic study is performed in order to 
determine the real carrier density profile induced on the graphene strip by the different gating pads. Then, this carrier density is employed to compute the electromagnetic behavior of the filter, as detailed in Section III.
The electrostatic problem has been solved using the commercial software ANSYS Maxwell, considering that the graphene strip is connected to the ground, and assigning adequate biasing voltages to the different gating pads.
These voltages are initially computed with Eq. \ref{eq: chemical vdc} to provide the ideal values of chemical potential. The gating pads are placed
  $25$~nm below the graphene strip  and $35$~nm apart from adjacent pads ($t = 25$~nm and $d = 35$~nm in Fig. \ref{fig: struct graph}), with a pad thickness of $50$~nm.
This approach allows to obtain the charge distribution at the graphene-substrate interface, $\rho(z)$, which in turn permits computing the real distribution of chemical potential along the strip as \cite{hanson2013softboundary}

\begin{equation}\label{eq: muc formula}
  \mu_c (z) = \frac{\hbar v_F}{e} \sqrt{\frac{\pi \rho(z)}{e}}.
\end{equation}

Once the chemical potential is known, the complex-valued conductivity at each point of the strip is computed with \eqref{eq:conductivity}.
Due to the finite distance between pads and the effect of fringing DC fields, using the initial voltages computed with Eq. \ref{eq: chemical vdc} may result in a filter with a shifted frequency response, requiring an additional optimization step. Tables \ref{tab: pads 1}-\ref{tab: pads 2} show the initial and optimized values of voltage assuming no previous chemical doping.
Figs. \ref{fig: muc and filters a}-\ref{fig: muc and filters b} depict the chemical potential profile along the strip computed with this approach for the filters proposed in the previous section.

Note that filters with lower gate voltages can be easily designed by setting a more relaxed initial specification, specifically by increasing the electrical length of the transmission line sections ($\theta_c$) in the synthesis of the polynomials. However, this comes at the expense of a worse spurious free range in the final filter.

Once the electrostatic problem has been solved, we analyze the EM response of the proposed structures taking into account the presence of the gating pads, which can be modelled at THz frequencies as a dielectric with permittivity $\varepsilon_r\approx 3$ \cite{esquius2014sinusoidallyleaky,pryputniewicz2002mems}. The inclusion of the gating pads in the dynamic analysis barely affects the EM response of the filters, because they are electrically very thin and with relative permittivity very similar to the background substrate.
Figs. \ref{fig: muc and filters c}-\ref{fig: muc and filters d} show the frequency response of these filters, computed via full-wave simulations and with the the scaling law, obtaining again good accuracy while requiring negligible computational resources.
This study further confirms the robustness and usefulness of the proposed analysis
technique, since performing full-wave analysis of continuously varying
conductivity profiles
is a tedious and time-consuming process.
The overall performance of both filters remains satisfactory, despite the deterioration of the in-band reflection profile.
This effect is caused by the smooth transitions in the spatial distribution of chemical potential, not accounted for in prototype network, and is more severe in the $9^{th}$ degree filter due to very strong and narrow variation of conductivity
in the central sections of the filter.
This known limitation of stepped impedance low-pass filters \cite{Cameron07} could be overcome by implementing more complex circuits that use impedance inverters or lumped elements to account for higher order effects at the junctions.
%
%
%
\subsection{Influence of Graphene losses}
\begin{figure}\centering
\subfloat[]{\label{fig: filters rs a}
\includegraphics[width=0.5\columnwidth]{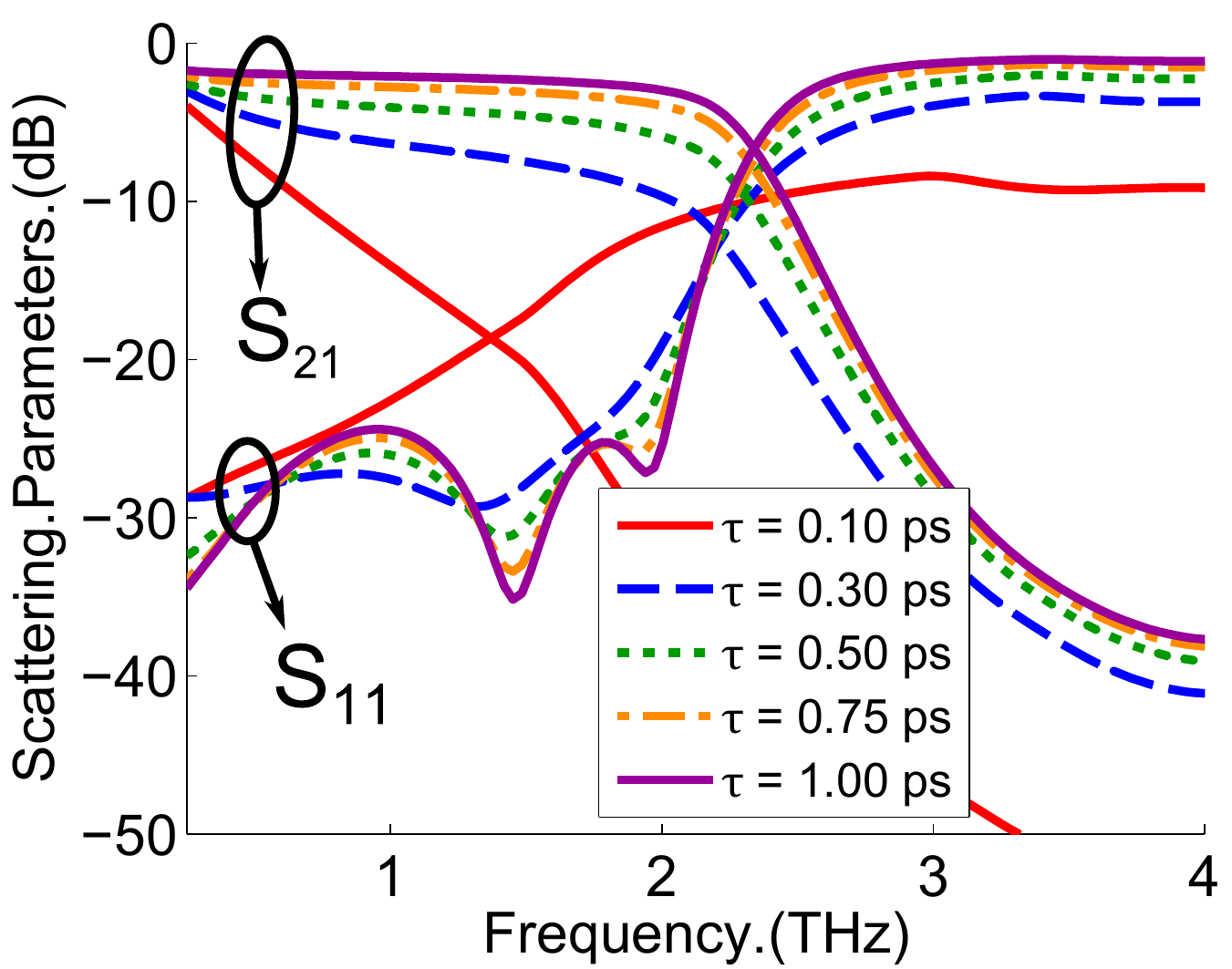}}
\subfloat[]{\label{fig: filters rs b}
\includegraphics[width=0.5\columnwidth]{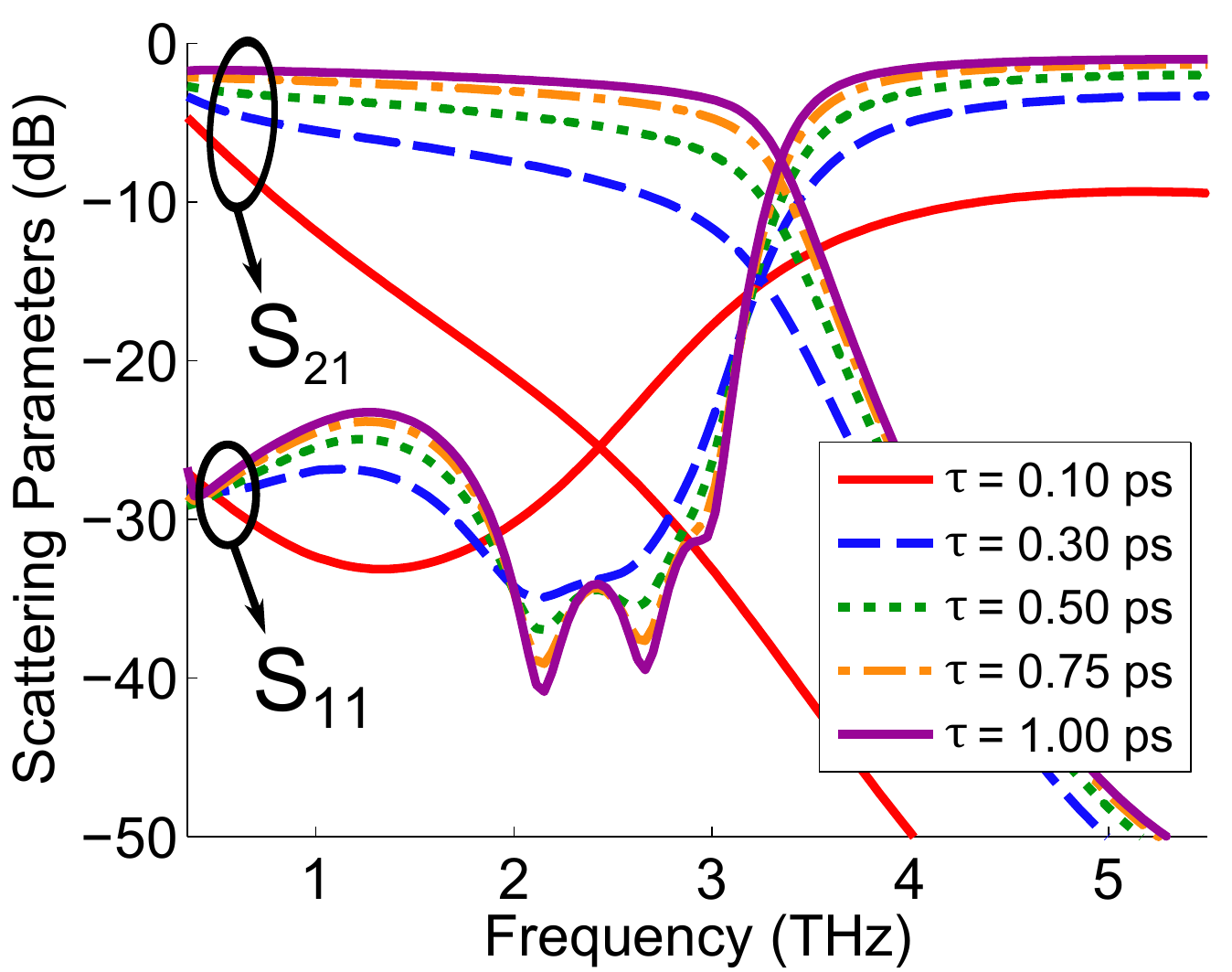}}
\caption{Influence of graphene's relaxation time in the frequency response of the (a) $7^{th}$ and (b) $9^{th}$ degree filters.}
\label{fig: filters rs}
\end{figure}

The presence of potentially high losses is an important factor to take into account while evaluating the performance of the proposed filters. To assess this point, the $7^{th}$ and $9^{th}$ degree filters have been analysed assuming different values of graphene relaxation time ($\tau$). Figs. \ref{fig: filters rs a}-\ref{fig: filters rs a} show the frequency response of both filters for values of $\tau$ ranging from $0.1$~ps to $1$~ps.
It is observed that graphene's relaxation time strongly affects the insertion loss of the filter and the sharpness of the transition between the pass-band and the rejected band. For relaxation times as low as $\tau = 0.1$~ps, losses are too high for the filter to present practical utility, whereas values nearing $0.5$~ps and above show very good performance.
Importantly, values of electron mobility in graphene on a boron nitride substrate
of up to $40000$ $cm^2/Vs$,
which corresponds to $\tau \approx 0.8$~ps, have been experimentally
observed at room temperature \cite{dean2010boron}. This measured value confirms that the proposed filters are of practical value using state of the art graphene technology.

\section{Conclusion}

We have proposed and analyzed graphene-based plasmonic low-pass filters in the THz band. An efficient synthesis method has been presented, based on classical microwave filter theory adapted to graphene SPPs through a recently introduced electrostatic scaling law. Several results, verified with full-wave simulations, showcase the filtering capabilities of the proposed structure and its
good performance compared to the alternatives in this frequency range, 
particularly in terms of miniaturization and reconfigurability. Through adequate use of electrostatic gating, the cut-off frequency of the device can be continuously tuned over a wide frequency range, resulting in a degree of tunability not achievable with any other existing technology in the THz band.
We envision that the proposed reconfigurable filtering structures may be useful in future all-integrated graphene plasmonic THz devices, sensors and communication systems.

\section*{Acknowledgment}
This work was supported by the Swiss National Science Foundation (SNSF) under grant $133583$ and by the EU FP7 Marie-Curie IEF grant Marconi, with ref. $300966$, Spanish Ministry of Education under grant TEC2013-47037-C5-5-R, and
European Feder Fundings. The authors wish to thank Mr. A. Manjavacas (CSIC, Spain) for fruitful discussions.


\end{document}